\documentclass[aip, apl, reprint, amsmath,amssymb,superscriptaddress]{revtex4-1}
\usepackage{graphicx}
\usepackage{hyperref}
\usepackage[normalem]{ulem}
\usepackage{xcolor}
\newcommand*{\citen}[1]{%
  \begingroup
    \romannumeral-`\x 
    \setcitestyle{numbers}%
    \cite{#1}%
  \endgroup   
}

\begin{document}

\title{Observation of Coherently Coupled Cation Spin Dynamics in an Insulating Ferrimagnetic Oxide}

\author{C.~Klewe} 
\email{cklewe@lbl.gov}
\affiliation{ 
Advanced Light Source, Lawrence Berkeley National Laboratory, Berkeley, CA, USA 
}%

\author{P.~Shafer}
\affiliation{ 
Advanced Light Source, Lawrence Berkeley National Laboratory, Berkeley, CA, USA }%

\author{J.~E.~Shoup}
\affiliation{Department of Physics, University of South Florida, Tampa, FL, USA }%

\author{C.~Kons}
\affiliation{Department of Physics, University of South Florida, Tampa, FL, USA }%

\author{Y.~Pogoryelov}
\affiliation{Department of Physics and Astronomy, Molecular and Condensed Matter Physics, Uppsala University, Uppsala, Sweden }%

\author{R.~Knut}
\affiliation{Department of Physics and Astronomy, Molecular and Condensed Matter Physics, Uppsala University, Uppsala, Sweden }%

\author{B.~A.~Gray}
\affiliation{Materials and Manufacturing Directorate, Air Force Research Lab, Wright Patterson Air Force Base, OH, USA}%

\author{H.-M.~Jeon}
\affiliation{KBR, Beavercreek, OH, USA }%

\author{B.~M.~Howe}
\affiliation{Materials and Manufacturing Directorate, Air Force Research Lab, Wright Patterson Air Force Base, OH, USA}%

\author{O.~Karis}
\affiliation{Department of Physics and Astronomy, Molecular and Condensed Matter Physics, Uppsala University, Uppsala, Sweden }%

\author{Y.~Suzuki}
\affiliation{Geballe Laboratory for Advanced Materials, Stanford University, Stanford, CA, USA }%
\affiliation{Department of Applied Physics, Stanford University, Stanford, CA, USA }%

\author{E.~Arenholz}
\affiliation{ 
Advanced Light Source, Lawrence Berkeley National Laboratory, Berkeley, CA, USA 
}%

\author{D.~A.~Arena}
\email{darena@usf.edu}
\affiliation{Department of Physics, University of South Florida, Tampa, FL, USA }%

\author{S.~Emori}
\email{semori@vt.edu}
\affiliation{Geballe Laboratory for Advanced Materials, Stanford University, Stanford, CA, USA }%
\affiliation{ 
Department of Physics, Virginia Tech, Blacksburg, VA, USA
}%

\date{9 January 2023}

\begin{abstract}
Many technologically useful magnetic oxides are ferrimagnetic insulators, which consist of chemically distinct cations. Here, we examine the spin dynamics of different magnetic cations in ferrimagnetic NiZnAl-ferrite (Ni$_{0.65}$Zn$_{0.35}$Al$_{0.8}$Fe$_{1.2}$O$_4$) under continuous microwave excitation. Specifically, we employ time-resolved x-ray ferromagnetic resonance to separately probe Fe$^{2+/3+}$ and Ni$^{2+}$ cations on different sublattice sites. Our results show that the precessing cation moments retain a rigid, collinear configuration to within $\approx$2$^\circ$. Moreover, the effective spin relaxation is identical to within $<$10\% for all magnetic cations in the ferrite. We thus validate the oft-assumed ``ferromagnetic-like'' dynamics in resonantly driven ferrimagnetic oxides, where the magnetic moments from different cations precess as a coherent, collective magnetization.  

\end{abstract}

\maketitle

Magnetic insulators are essential materials for computing and communications devices that rely on spin transport without net charge transport~\cite{Brataas2020, Chumak2015}. Most room-temperature magnetic insulators possess antiferromagnetically coupled sublattices~\cite{Jungwirth2016, Gomonay2017, Baltz2018, Emori2021, Kim2021}.  Many are true antiferromagnets with prospects for ultrafast spintronic devices~\cite{Jungwirth2016, Gomonay2017, Baltz2018}. Yet, challenges remain in controlling and probing the magnetic states of antiferromagnets~\cite{Gray2019, Meer2021, Cogulu2021}. For practical applications, perhaps more promising insulators are ferrimagnetic oxides~\cite{Harris2012, Emori2021, Kim2021} – such as iron garnets and spinel ferrites – that possess unequal sublattices incorporating different cations. The magnetization state in such ferrimagnets can be straightforwardly controlled and probed by well-established methods, i.e., via applied magnetic fields and spin currents~\cite{Kim2021}. Further, the properties of ferrimagnetic oxides (e.g., damping, anisotropy) can be engineered by deliberately selecting the cations occupying each sublattice~\cite{Harris2012,Emori2021, Dionne2011}. 

Most studies to date have effectively treated ferrimagnetic oxides as \emph{ferro}magnets: the cation magnetic moments are presumed to remain collinear and coherent while they are excited, such that they behave as one ``net'' magnetization (i.e., the vector sum of the cation moments). However, it is reasonable to question how much these cation moments can deviate from the ferromagnetic-like dynamics. Such deviations may be plausible, considering that the coupling among the cations may not be perfectly rigid or that different magnetic cations in the sublattices may exhibit different rates of spin relaxation (effective damping)~\cite{Harris2012, Dionne2011}. Indeed, a recent experimental study on biaxial yttrium iron garnet demonstrates peculiar spin-torque switching results~\cite{Zhou2020}, suggesting that ferrimagnetic oxides -- even with a large net magnetization -- could deviate from the expected ferromagnetic-like dynamics. Given the application potential and fundamental interest, it is timely to explore the dynamics of specific sublattices and cations in ferrimagnetic oxides. 

\begin{figure}[b]
	\centering
		\includegraphics[width=4.5cm]{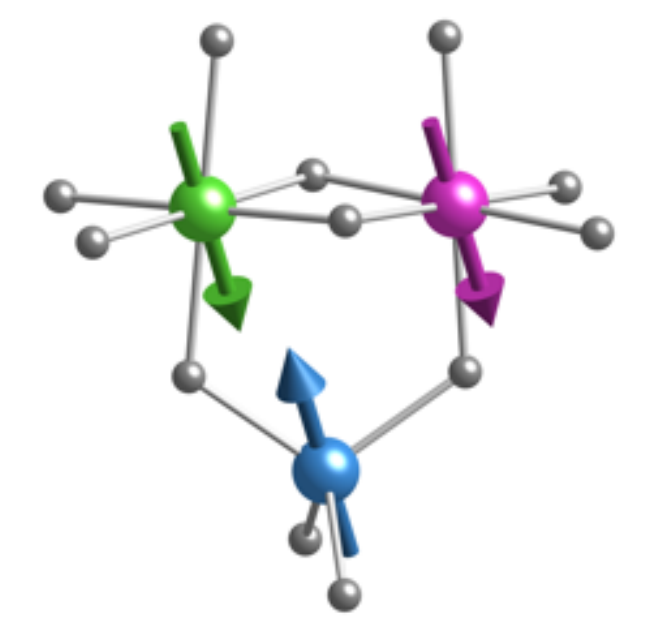}
	\caption{Schematic of a portion of the spinel structure, showing two cations (e.g., Fe$^{3+}_{O_h}$, Ni$^{2+}_{O_h}$) occupying the octahedrally-coordinated sublattice (green and purple) and a cation (e.g., Fe$^{3+}_{T_d}$) occupying the tetrahedrally-coordinated sublattice (blue). The gray spheres represent oxygen anions.}	
	\label{fig:sublattices}
\end{figure}

In this Letter, we present unprecedented experimental insight into resonant spin dynamics in a multi-cation ferrimagnetic oxide. Specifically, we investigate sublattice- and cation-specific dynamics in NiZnAl-ferrite (Ni$_{0.65}$Zn$_{0.35}$Al$_{0.8}$Fe$_{1.2}$O$_4$), a spinel ferrimagnetic oxide with two magnetic sublattices [Fig.~\ref{fig:sublattices}]: (i) the tetrahedrally coordinated sublattice, $T_d$, predominantly consisting of Fe$^{3+}_{T_d}$ cations and (ii) the octahedrally coordinated sublattice, $O_h$, predominantly consisting of Fe$^{3+}_{O_h}$, Fe$^{2+}_{O_h}$, and Ni$^{2+}_{O_h}$ cations. 
We utilize x-ray ferromagnetic resonance (XFMR)~\cite{Boero2005,Arena2006, Arena:2007, Guan2007, ArenaRSI2009, Warnicke2015, Baker2016, Li2016c, Li2019d, Dabrowski2020, Emori2020b, Klewe2020, Klewe2022}, which leverages x-ray magnetic circular dichroism (XMCD) that is sensitive to chemical elements, site coordination, and valence states. With this XFMR technique, we detect the precessional phase and amplitude for $\emph{each}$ magnetic cation species. 

Our cation-specific XFMR measurements are further augmented by the following attributes of NiZnAl-ferrite. 
First, the NiZnAl-ferrite film exhibits about two orders of magnitude lower magnetic damping than the ferrite in an earlier XFMR study~\cite{Warnicke2015}, yielding a far greater signal-to-noise ratio in XFMR measurements. This permits comprehensive measurements at multiple applied magnetic fields, which allow precise quantification of the precessional phase lags among the cation species. 
Second, NiZnAl-ferrite is an intriguing test-bed for exploring whether the excited magnetic cations retain collinear coupling. The nonmagnetic Zn$^{2+}$ and Al$^{3+}$ cations dilute the magnetic exchange coupling in NiZnAl-ferrite~\cite{Wilber1983}, as evidenced by a modest Curie temperature of $\approx 450$ K \cite{Emori2017}, such that the magnetic Fe$^{2+/3+}$ and Ni$^{2+}$ cations may not remain rigidly aligned.  
Lastly, with diverse magnetic cations in NiZnAl-ferrite, we address whether cations with different spin-orbit coupling can exhibit distinct spin relaxation~\cite{Dionne2011} by quantifying the FMR linewidths and precessional cone angles for the different cations. 
Taken together, we are able to probe -- with high precision -- the possible deviation from the oft-assumed ferromagnetic-like dynamics in the ferrimagnetic oxide.

Our study focuses on a 23-nm thick epitaxial NiZnAl-ferrite film grown on (001) oriented, isostructural MgAl$_2$O$_4$ substrates by pulsed laser deposition~\cite{Emori2017}. 
The NiZnAl-ferrite film, magnetized along the [100] direction, was probed at room temperature with a circularly polarized x-ray beam at Beamline 4.0.2 at the Advanced Light Source (ALS), Lawrence Berkeley National Laboratory. 
The XFMR measurements follow a pump-probe method: the RF excitation (4-GHz pump) is synchronized to a higher harmonic of the x-ray pulse frequency (500-MHz probe), and the transverse component of the precessing magnetization is probed stroboscopically.
A variable delay between the RF pump signal and the timing of the x-ray pulses enables mapping of the complete magnetization precession cycle.
A photodiode mounted behind the sample collects the luminescence yield from the subjacent MgAl$_2$O$_4$ substrate. The luminescence yield detection enables the investigation of high-quality epitaxial films on single-crystal substrates ~\cite{Baker2016, Li2019d, Dabrowski2020,  Klewe2022}. This is in contrast to transmission detection that is limited to polycrystalline films on thin membrane substrates~\cite{Guan2007, Warnicke2015}. A more detailed description of the XFMR setup is provided in Refs.~\citen{Klewe2020} and \citen{Klewe2022}.

\begin{figure}[t]
	\centering
		\includegraphics[width=8.6cm]{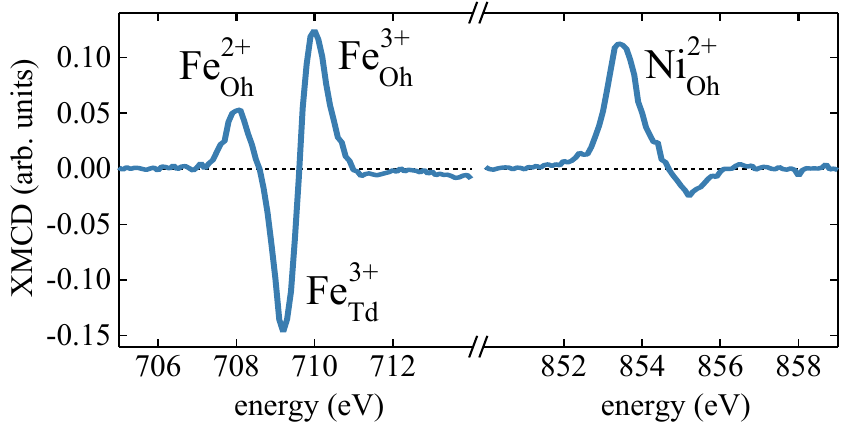}
	\caption{Static XMCD spectra taken at the $L_3$ edge of Fe and Ni. The characteristic peaks are attributed to the corresponding cation valence states and sublattice site occupation.}	
	\label{fig:static_XMCD}
\end{figure}

By tuning the photon energy to the element- and coordination-specific features in the static XMCD spectra, we are able to probe the magnetism of different elements, valence states, and sublattice sites \emph{individually}.
Static XMCD spectra at the $L_{3}$ edge of Fe and Ni are shown in Fig. \ref{fig:static_XMCD}.
The spectra show pronounced peaks from different cations on the $O_h$ and $T_d$ sublattices. 
While an XMCD spectrum is generally a complicated superposition of different coordinations and valence states, the three distinct peaks in the Fe  $L_{3}$ spectrum at $708.0\,$eV, $709.2\,$eV, and $710.0\,$eV are attributed to Fe$^{2+}_{O_h}$, Fe$^{3+}_{T_d}$, and Fe$^{3+}_{O_h}$, respectively, to a good approximation~\cite{Pattrick2002, Hoppe2015}. 
The opposite polarities of the Fe$^{3+}_{T_d}$ and Fe$^{2+/3+}_{O_h}$ peaks reflect the antiferromagnetic coupling between the $T_d$ and $O_h$ sublattices at static equilibrium.
Ni$^{2+}$ cations predominantly occupy the $O_h$ sublattice~\cite{Pattrick2002, Emori2017}, such that the XMCD peak at $853.5\,$eV is assigned to Ni$^{2+}_{O_h}$.

\begin{figure}[t]
	\centering
		\includegraphics[width=8.6cm]{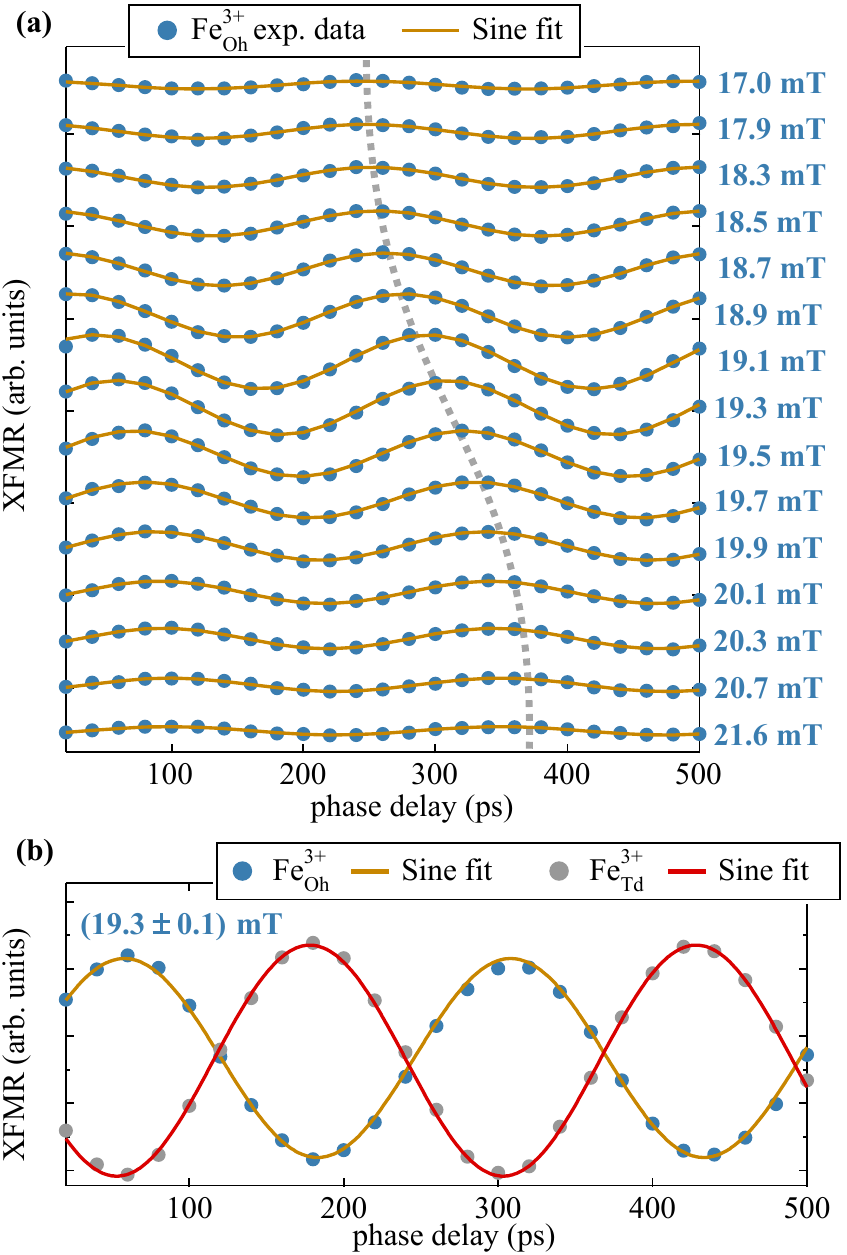}
	\caption{(a) Bias field resolved phase delay scans at the resonant core level excitation energy of Fe$^{3+}_{O_h}$ ($710.0\,$eV). The dashed curve highlights the characteristic shift across the resonance. (b) Comparison between delay scans of Fe$^{3+}_{T_d}$ ($709.2\,$eV) and Fe$^{3+}_{O_h}$ ($710.0\,$eV) cations taken at $19.3\,$mT.}	
	\label{fig:Fe3+Oh_delays}
\end{figure}

XFMR measurements were carried out at the photon energies specific to the cations found above. 
For each cation, we performed phase delay scans to map out the precession at different field values across the resonance field $\mu_0 H_\mathrm{res}$. 
Figure \ref{fig:Fe3+Oh_delays}(a) displays a set of phase delay scans taken at a photon energy of $710.0\,$eV, corresponding to Fe$^{3+}_{O_h}$. Each scan was taken at a fixed bias field between $17.0\,$mT and $21.6\,$mT. 
The phase delay scans exhibit pronounced oscillations with a periodicity of $250\,$ps in accordance with the 4-GHz excitation. 
Figure~\ref{fig:Fe3+Oh_delays}(b) depicts delay scans for Fe$^{3+}_{T_d}$ ($709.2\,$eV) and Fe$^{3+}_{O_h}$ ($710.0\,$eV) taken at $\mu_0 H$ = 19.3 mT (center of the resonance curve). The opposite sign of the two oscillations indicates a phase shift of about $180^\circ$ between the two sublattices. The result in  Fig.~\ref{fig:Fe3+Oh_delays}(b) thus suggests that the moments of Fe$^{3+}_{T_d}$ ($709.2\,$eV) and Fe$^{3+}_{O_h}$ in NiZnAl-ferrite maintain an antiferromagnetic alignment during resonant precession. 

\begin{figure*}
    \centering
    \includegraphics[width=0.99\textwidth]{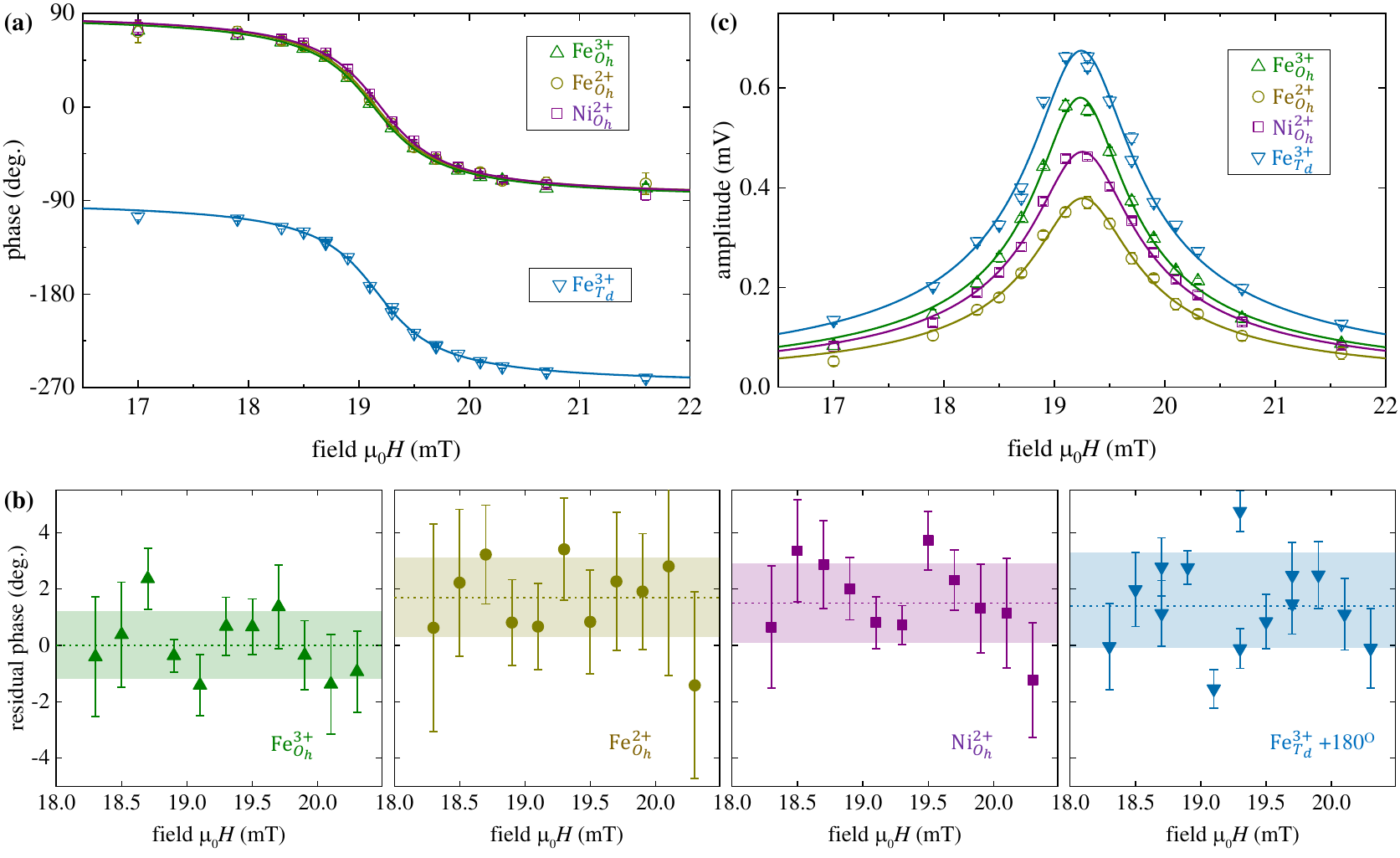}
    \caption{(a) Field dependence of the precessional phase for each magnetic cation species. The solid curve indicates the fit result with Eq.~\ref{eq:arctan}. (b) Residuals of the precessional phase (i.e., difference between the experimentally measured data and the $\arctan(\Delta H/(H-H_\mathrm{res}))$ part of the fit curve) in the vicinity of the resonance field. The dashed horizontal line indicates the phase lag $\phi_0$ relative to the precessional phase of Fe$^{3+}_{O_h}$. The shaded area indicates the standard deviation of the data points shown in each panel. (c) Field dependence of the precessional amplitude. The solid curve indicates the fit result with Eq.~\ref{eq:ampl}.}
    \label{fig:ampl_phase_comparison}
\end{figure*}

In the remainder of this Letter, we quantify the precessional phase and relaxation of each cation by analyzing our field-dependent XFMR results, summarized in Fig. \ref{fig:ampl_phase_comparison}.
Figure~\ref{fig:ampl_phase_comparison}(a) shows that all cations in NiZnAl-ferrite exhibit a characteristic $180^\circ$ phase reversal across the resonance of a damped harmonic oscillator. 
Quick visual inspection reveals that all $O_h$ cations are approximately in phase. Further, the $O_h$ and $T_d$ cations are approximately $180^\circ$ out of phase, as expected for the precession of antiferromagnetically coupled moments. 

To quantify the phase lag among the cations precisely, the field dependence of the precessional phase $\phi$ for each cation is modeled with
\begin{equation}
\label{eq:arctan}
\phi =\phi_\mathrm{0} + \arctan\left(\frac{\Delta H_\mathrm{hwhm}}{H-H_\mathrm{res}}\right),
\end{equation}
where $\phi_0$ is the baseline of the precessional phase (set to 0 for Fe$^{3+}_{O_h}$) and $\Delta H_\text{hwhm}$ is the half-width-at-half-maximum FMR linewidth. Equation~\ref{eq:arctan} is equivalent to the expressions in Refs.~\citen{Guan2007} and \citen{Li2016c} and valid when the effective magnetization (including the out-of-plane magnetic anisotropy) $\mu_0M_\mathrm{eff}\approx1$~T ~\cite{Emori2017} is much larger than the applied bias field $\mu_0H \approx 20$~mT. 
We quantify $H_\mathrm{res}$ and $H_\mathrm{hwhm}$ by simultaneously fitting the field dependence of the precessional phase $\phi$ [Eq.~\ref{eq:arctan}] and of the precessional amplitude $A$,
\begin{equation}
\label{eq:ampl}
A \propto \sqrt{\frac{\Delta {H_\mathrm{hwhm}}^2}{\Delta {H_\mathrm{hwhm}}^2+(H-H_\mathrm{res})^2}}.
\end{equation}  
To account for reduced sensitivity far from the resonance, the fits in Fig. \ref{fig:ampl_phase_comparison}(a,c) are weighted using the error bars from the sinusoidal fits of the phase delay scans (e.g., Fig.~\ref{fig:Fe3+Oh_delays}(b)). The results of the fitting are shown in Fig.~\ref{fig:ampl_phase_comparison}(a,c) and Table~\ref{tab:fit_results}. 

If the magnetic moments of the four cations were perfectly collinear, the phase lag should be $\phi_0 = 0$ for the $O_h$ cation species whereas $\phi_0 = 180^\circ$ for the $T_d$ cation species. 
Taking Fe$^{3+}_{O_h}$ as the reference, the results in  Table~\ref{tab:fit_results} show that $\phi_0$ deviates by $\approx$1.5$^\circ$ from the perfect collinear scenario. However, we caution that the uncertainty of $\phi_0$ in Table~\ref{tab:fit_results} is likely underestimated. Indeed, by examining the residuals of the fits displayed in Fig.~\ref{fig:ampl_phase_comparison}(b), we observe a scatter in the measured precessional phase of at least $\approx$2$^\circ$. It is sensible to conclude that the $O_h$ cations maintain a relative precessional phase lag of $(0\pm2)^\circ$, whereas the $O_h$ and $T_d$ cations maintain a phase lag of $(180\pm2)^\circ$. 
Even with the diluted exchange coupling from nonmagnetic Zn$^{2+}$ and Al$^{3+}$ cations, the magnetic Fe$^{2+/3+}$ and Ni$^{2+}$ cations retain a coherent, collinear alignment.

Reducing the experimental uncertainty to well below 2$^\circ$ would be extremely challenging.  
For each cation, a small drift in the beamline photon energy with respect to its XMCD peak (Fig. \ref{fig:static_XMCD}) might shift its apparent precession phase, due to an overlap in the cation specific XMCD features. For instance, considering that the difference between the Fe$^{3+}_{T_d}$ and Fe$^{3+}_{O_h}$ peaks is only $\approx$0.8 eV, an energy drift of $\approx$0.01 eV could cause a phase shift of $\approx$2$^\circ$. 
The nominal resolution of the electromagnet at $\approx$0.1 mT may also contribute to the scatter in the field dependence of XFMR phase.  
Moreover, the timing jitter of the master oscillator of up to $\approx$3 ps limits the time resolution of the phase delay scans. 
Taking all the above factors into account, the resolution of $\approx$2$^\circ$ in our present study is in fact at the practical limit.  

We now provide insight into the spin relaxation of each magnetic cation species by quantifying the cation-specific FMR linewidth $\Delta H_\mathrm{hwhm}$. 
In particular, we examine whether different spin relaxation emerges for magnetic cations with different strengths of spin-orbit coupling -- e.g., Fe$^{3+}$ with nominally zero orbital angular momentum vs Fe$^{2+}$ with likely nonzero orbital angular momentum~\cite{Dionne2011}. 
However, Fig.~\ref{fig:ampl_phase_comparison}(c) and Table~\ref{tab:fit_results} show that all magnetic cations in NiZnAl-ferrite exhibit essentially the same linewidth, $\mu_0 \Delta H_\text{hwhm}=(0.43 \pm 0.03)\,$mT, consistent with the value obtained from conventional FMR for NiZnAl-ferrite~\cite{Emori2017}. 
Our finding thus indicates that the exchange interaction in NiZnAl-ferrite leads to uniform spin relaxation across all magnetic cations. 

To further characterize cation-specific spin relaxation, we quantify the precessional cone angle $\theta_\mathrm{cone}$ of each magnetic cation species. Specifically, $\theta_\mathrm{cone}$ is obtained from the amplitudes of the XFMR signal $I_\text{XFMR}$ and XMCD peak $I_\text{XMCD}$ via
\begin{equation}
\label{eq:coneangle}
\theta_\mathrm{cone} = 2 \arcsin\left(\frac{I_\text{XFMR}}{I_\text{XMCD}}\right).
\end{equation}
We find that all cation moments precess with a cone angle of $\theta \approx 1.0-1.1^\circ$. 
Our results confirm the exchange interaction in NiZnAl-ferrite is strong enough to lock all magnetic cations at the same relaxation rate, as evidenced by the invariance of the linewidth and precessional cone angle to within $\lesssim$10\%.  

Our finding is distinct from recent work on a ferrimagnetic DyCo alloy, showing different damping parameters for two magnetic sublattices after femtosecond-laser-induced demagnetization~\cite{Abrudan2021}. 
In Ref.~\citen{Abrudan2021}, the laser pulse produces a highly nonequilibrium distribution of spins, which in turn quenches the exchange interactions, and the two ferrimagnetic sublattices are free to relax quasi-independently of each other. In this case, the rare-earth Dy sublattice with stronger spin-orbit coupling exhibits a higher damping parameter than the transition-metal Co sublattice. 
By contrast, the magnetic moments in our experiment are forced to oscillate by continuous microwave excitation, yet remain at near-equilibrium across the entire resonance curve. The near-equilibrium forced oscillations -- in concert with the exchange interactions -- favor the rigid, coherent coupling among the magnetic cations and sublattices.

\begin{table}[t!]
\centering
\caption{Resonance field $H_\mathrm{res}$, linewidth $\Delta H_\mathrm{hwhm}$, relative precessional phase lag $\phi_0$, and precessional cone angle $\theta_\mathrm{cone}$ for each magnetic cation species, as derived from fitting the field dependence of the precessional phase [Eq.~\ref{eq:arctan}] and amplitude [Eq.~\ref{eq:ampl}].}
\begin{tabular}{@{}lrrrr@{}}
\hline 
Cation & $\mu_0  H_\text{res}$ & $\mu_0 \Delta H_\text{hwhm}$ & $\phi_0$ & $\theta_\mathrm{cone}$ \\ 
 & (mT) & (mT) & (degree) & (degree)\\
\hline 
Fe$^{3+}_{O_h}$ & $19.21 \pm 0.03$  & $0.43 \pm 0.03$ & 0 & $1.0 \pm 0.1$  \\[1ex]
Fe$^{2+}_{O_h}$ & $19.22 \pm 0.03$  & $0.44 \pm 0.03$ & $1.7 \pm 1.2$ & $1.0  \pm 0.1$ \\[1ex]
Ni$^{2+}_{O_h}$ & $19.23 \pm 0.02$  & $0.43 \pm 0.02$ & $1.5 \pm 1.2$ & $1.1  \pm 0.1$  \\[1ex]
Fe$^{3+}_{T_d}$  & $19.22 \pm 0.03$  & $0.43 \pm 0.03$ & $-178.5 \pm 1.4$ & $1.1 \pm 0.1$  \\[1ex]     
\hline
\label{tab:fit_results}
\end{tabular}
\end{table}

In summary, we have investigated time-resolved, cation-specific resonant magnetic prcession at room temperature in an epitaxial thin film of NiZnAl-ferrite, a spinel-structure insulating ferrimagnetic oxide. The low damping of this ferrite film yields a large XFMR signal-to-noise ratio, allowing us to resolve precessional dynamics with high precision. 
In particular, we have obtained two key findings. First, the magnetic cations retain a coherent, collinear configuration, to within an uncertainty in the precessional phase of $\approx$2$^\circ$. Second, the strongly coupled magnetic cations experience the same magnitude of spin relaxation, to within an uncertainty of $\lesssim$10\%. 
Thus, the oft-assumed ``ferromagnet-like'' dynamics remain robust in the ferrimagnetic oxide, even with high contents of nonmagnetic Zn$^{2+}$ and Al$^{3+}$ cations that reduce the exchange stiffness. 
We emphasize that our conclusion is specific to the resonant dynamics under a \emph{continuous-wave} excitation. Future time-resolved XMCD measurements may resolve cation-specific dynamics in ferrimagnetic oxides driven by sub-nanosecond \emph{pulses}, e.g., of electric-current-induced torques, with potential implications for ultrafast device technologies. 

\section*{Acknowledgements}
C.K. acknowledges financial support by the Alexander von Humboldt foundation. S.E. and Y.S. were funded by the Vannevar Bush Faculty Fellowship of the Department of Defense under Contract No. N00014-15-1-0045. Work by S.E. was also supported in part by the Air Force Office of Scientific Research under Grant No. FA9550-21-1-0365. Y.S. was also funded by the Air Force Office of Scientific Research  under Grant No. FA9550-20-1-0293. D.A.A. acknowledges the support of the National Science Foundation under Grant No. ECCS-1952957 and also the USF Nexus Initiative and the Swedish Fulbright Commission.  The Advanced Light Source is supported by the Director, Office of Science, Office of Basic Energy Sciences, of the U.S. Department of Energy under Contract No. DE-AC02-05CH11231.

\section*{Data Availability}
The data that support the findings of this study are available from the corresponding author upon reasonable request.

\end{document}